\documentclass[aps,pra,showpacs,amsmath,amssymb,amsfonts,superscriptaddress,lengthcheck,twocolumn,longbibliography]{revtex4-1}

\usepackage{dcolumn}    
\usepackage{bm} 
\usepackage{graphicx} 
\usepackage{latexsym}
\usepackage{array}      
\usepackage{epsfig}
\usepackage{epsf}
\usepackage{txfonts}
\usepackage{color}
\usepackage{bbm}
\usepackage{subfigure}
\usepackage{verbatim}
\usepackage[colorlinks=true,linkcolor=blue,urlcolor=blue,citecolor=blue,pdfusetitle]{hyperref}
\usepackage{hyperref}

\newcommand{\bra}[1]{\left\langle #1\right|}
\newcommand{\ket}[1]{\left|#1\right\rangle}

\newcommand{\etal}{\textit{et al. }}

\newcommand{\co}[1]{\cos{\left(#1\right)}}
\newcommand{\si}[1]{\sin{\left(#1\right)}}

\newcommand{\bla}{bla\\bla\\bla\\bla\\bla}

\newcommand{\rev}[1]{{\color{black}#1}}

\newcommand{\nbar}{\overline{n}}

\begin{document}
\title{Precision thermometry and the quantum speed limit}
\author{Steve Campbell}
\email{steve.campbell@mi.infn.it}
\affiliation{Istituto Nazionale di Fisica Nucleare, Sezione di Milano, \& Dipartimento di Fisica, Universit{\`a} degli Studi di Milano, Via Celoria 16, 20133 Milan, Italy}
\author{Marco G. Genoni}
\email{marco.genoni@fisica.unimi.it}
\affiliation{Quantum Technology Lab, Dipartimento di Fisica, Universit\`a degli Studi di Milano, 20133 Milano, Italy}
\author{Sebastian Deffner}
\email{deffner@umbc.edu}
\affiliation{Department of Physics, University of Maryland Baltimore County, Baltimore, MD 21250, USA}

\begin{abstract}
We assess precision thermometry for an arbitrary single quantum system. For a $d$-dimensional harmonic system we show that the gap sets a single temperature that can be optimally estimated. Furthermore, we establish a simple linear relationship between the gap and this temperature, and show that the precision exhibits a quadratic relationship. We extend our analysis to explore systems with arbitrary spectra, showing that exploiting anharmonicity and degeneracy can greatly enhance the precision of thermometry. Finally, we critically assess the dynamical features of two thermometry protocols for a two level system. By calculating the quantum speed limit we find that, despite the gap fixing a preferred temperature to probe, there is no evidence of this emerging in the dynamical features.
\end{abstract}
\date{\today}
\maketitle

\section{Introduction}

The zeroth law of thermodynamics asserts \cite{Callen1985}:
\begin{quote}
\textit{If two systems are in thermal equilibrium with a third system, then they are in thermal equilibrium with each other.}
\end{quote}
Different thermal equilibria are labeled by a single parameter -- \emph{temperature} --, which quantifies the subjective notion of hot and cold. Whereas the temperature of a classical system is one of the best understood and most commonly used physical quantities, assigning a meaningful and unique temperature to quantum systems is \textit{a priori} a significantly harder task \cite{Gemmer2009}. \rev{Indeed, generally} the temperature of quantum systems is neither a classical nor a quantum observable. Thus, one has to resort to quantum estimation techniques \cite{HelstromBook,ParisIJQI2009} to derive the ultimate limits on its determination. To this end, recent years have witnessed intense efforts in the design of `optimal quantum thermometers' and in accurately determining the temperature of a variety of quantum systems \cite{Stace2010,Monras2011,LuisPRL,BrunelliPRA2011,BrunelliPRA2012,Sabin2014,Jevtic2015,Guo2015,MehboudiNJP2015,MehboudiPRA2016, ParisJPA2016,DePasqualeNatComm2016,DePasqualePRA2017, DePasquale2017, CampbellNJP2017,Mancino2017, Oliveira2015}. 

However, several important issues have remained unsatisfactorily addressed: (i) what is the effect of the energy spectrum of a quantum system on the precision with which its temperature can be estimated? and (ii) how do fundamental quantum principles such as the indeterminacy relations affect the timescales over which the temperature of quantum systems can be estimated? In the following we analyze both questions with the help of analytically solvable case studies.

To begin, we consider the estimation of temperature via minimal thermometers comprised of individual quantum probes that are already at thermal equilibrium. Following from and generalizing the approach taken by Correa \etal~\cite{LuisPRL}, we start by discussing thermometry via quantum systems characterized by harmonic (equally gapped) spectra and discuss the role of the level spacing $\Delta$. We show that qualitatively identical results are obtained regardless of the dimensionality of the probe: for a given spacing $\Delta$, there is a single optimal temperature $T_\text{max}$ corresponding to a maximum in the estimation precision attainable, and precise functional relationships between these quantities can be readily obtained. We then consider quantum probes endowed with anharmonic (non-equally gapped) spectra and allow for degeneracy. We find, in line with Ref.~\cite{LuisPRL}, that degeneracy can in fact increase the estimation precision obtainable with a finite-dimensional quantum system. Furthermore we show how these two properties, anharmonicity and degeneracy, can be harnessed to allow for high precision estimation of more than a single temperature, as witnessed by the emergence of multiple peaks in the quantum Fisher information. 

The close connection between the energy spectrum and the optimal estimation of a quantum system's temperature also hints at a relation with the quantum speed limit (QSL), see recent reviews \cite{Frey2016QIP,Deffner2017JPA} and references therein. The QSL determines the shortest timescale over which quantum systems can evolve, and it can be interpreted as a consequence of Heisenberg's indeterminacy principle for energy and time. It governs the maximal precision of estimating the energy \cite{LloydPRL2006,LloydPRL2012}, and therefore one naturally would expect a close relationship between the QSL and temperature.  For instance, in quantum statistical physics it has proven mathematically useful to interpret temperature as a characteristic time-scale, albeit in imaginary time \cite{Nieto1995}.

We critically assess such a possible relation between the QSL and the optimal estimation of temperature by allowing a harmonic probe system to interact with finite and infinite dimensional environments whose temperature we are interested in determining. Interestingly, we find that the dynamics does not carry information about the optimal precision of thermometry, i.e. although there is a single optimal temperature dictated by the gap, no clear footprint of this is reflected in the dynamical features. In particular, despite the fact the QSL correctly characterizes the relaxation and thermalization dynamics, we find that the optimal estimation of temperature is independent of the maximal speed, and the corresponding QSL time, with which the quantum system evolves.

This work is organized as follows: in Sec.~\ref{s:harmonic} we introduce some basic notions of quantum estimation theory and present the results for thermometry with probes characterized by harmonic spectra. In Sec.~\ref{s:anharmonic} we discuss the possible enhancement that can be obtained for thermometry via probes with anharmonic and highly-degenerated spectra. In Sec.~\ref{s:QSL} we introduce the basic concept of the QSL and discuss its relationship with thermometry by considering two alternative dynamical schemes: firstly, we let the probe thermalize, in the usual quantum open system scenario, via a Markovian master equation. Secondly, we consider a simple toy-model where the probe interacts unitarily with a finite dimensional thermal environment. We conclude with some final remarks in Sec.~\ref{s:conclusions}.

\section{Thermometry for Harmonic Spectra} \label{s:harmonic}
Here we assess the effect that the energy level spacing and dimensionality have on precision thermometry for systems with harmonic spectra. In particular we first recap and elucidate some known results for the limiting cases of two- and infinite-dimensional systems \rev{\cite{LuisPRL,ParisJPA2016}}, highlighting the clear role that the single characteristic energy spacing plays, before extending our results to arbitrary dimensions. To this end, we will assume the system is already at thermal equilibrium, and therefore in a canonical Gibbs state, \rev{$\varrho_T = e^{-H/T}/\mathcal{Z} = \sum_n p_n |E_n\rangle\langle E_n|$, where $\{|E_n\rangle \}$ are the eigenstates of the Hamiltonian $H$, $\{p_n\}$} the corresponding populations for the thermal states, \rev{and $\mathcal{Z}=\text{Tr}\left[ e^{-H/T} \right]$ the associated partition function}. 

The precision with which temperature, $T$, can be estimated is bounded by the quantum Cram\'er-Rao bound
\begin{equation}
\text{Var}(T) \geq \frac{1}{M \mathcal{H}(T)},
\end{equation}
where $M$ is the number of measurements performed and $\mathcal{H}$ is the quantum Fisher information (QFI) \cite{HelstromBook,ParisIJQI2009}. The QFI depends only on the family of quantum thermal states $\varrho_T$ and, evidently, larger values of QFI correspond to a more accurate estimation of the temperature. In fact the QFI can be interpreted as the distance between two thermal states whose temperature differs by an infinitesimal variation, in formula
\begin{align}
\mathcal{H}(T) = 
8 \lim_{\delta T \to 0} \frac{1 - F[\varrho_T,\varrho_{T+\delta T}]}{\delta T^2} \,,
\end{align}
where $F[\varrho,\sigma] = \hbox{Tr}[\sqrt{\sqrt{\sigma}\varrho \sqrt{\sigma}}]$ denotes the fidelity
between two quantum states.

\begin{figure}[t]
{\bf (a)}\\
\includegraphics[width=0.8\columnwidth]{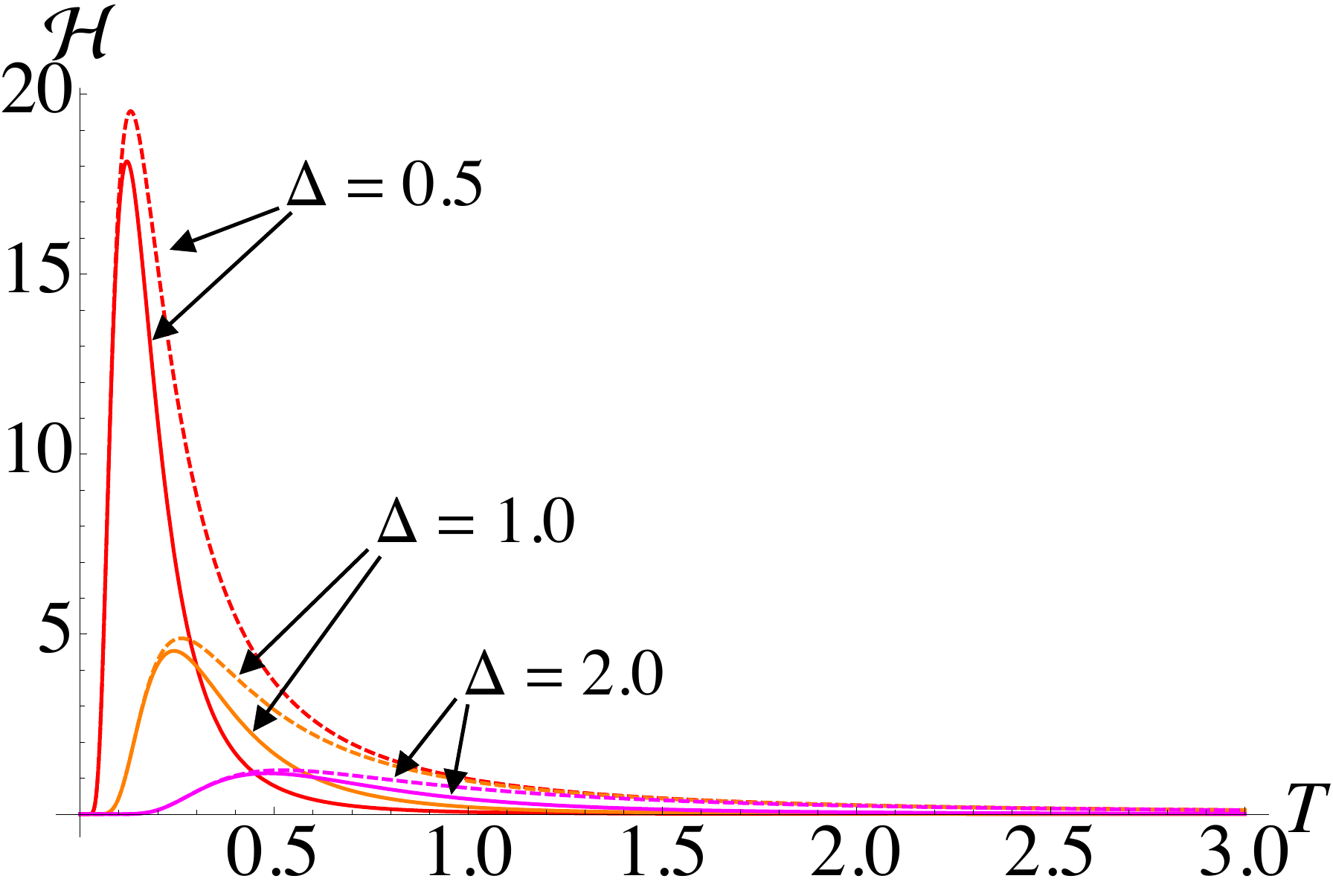}\\
{\bf (b)}\\
\includegraphics[width=0.8\columnwidth]{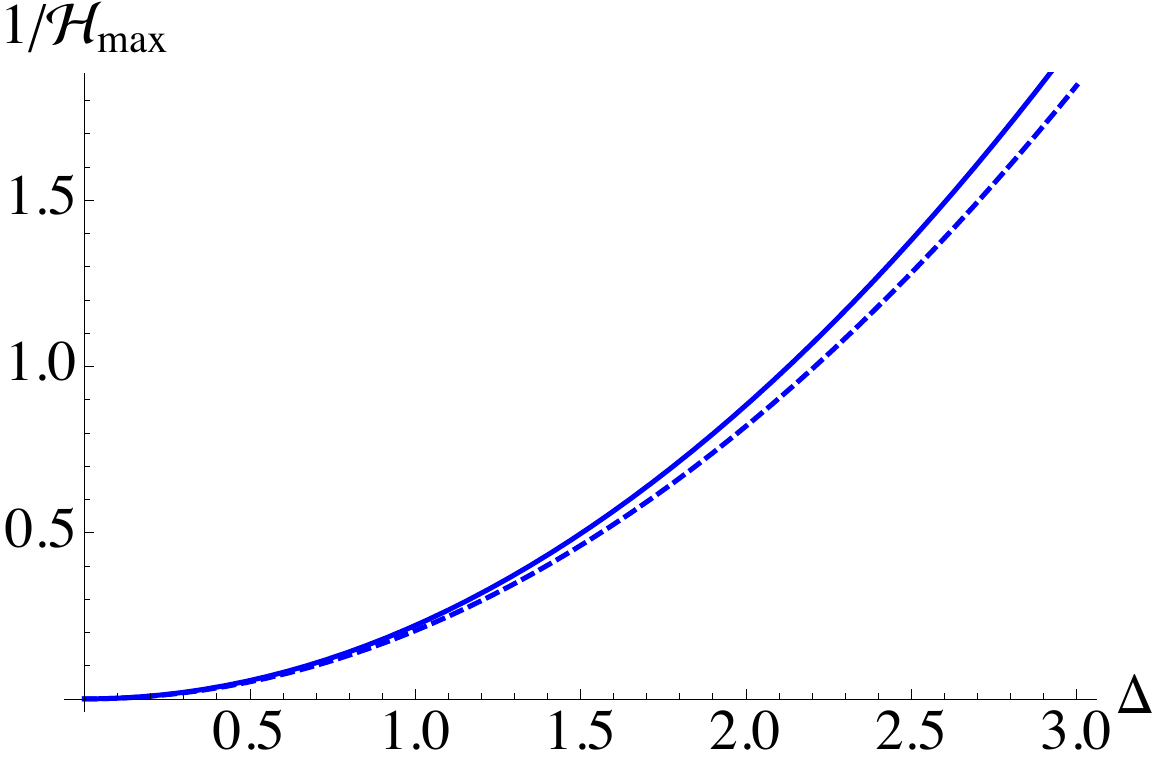}\\
 {\bf (c)}\\
\includegraphics[width=0.8\columnwidth]{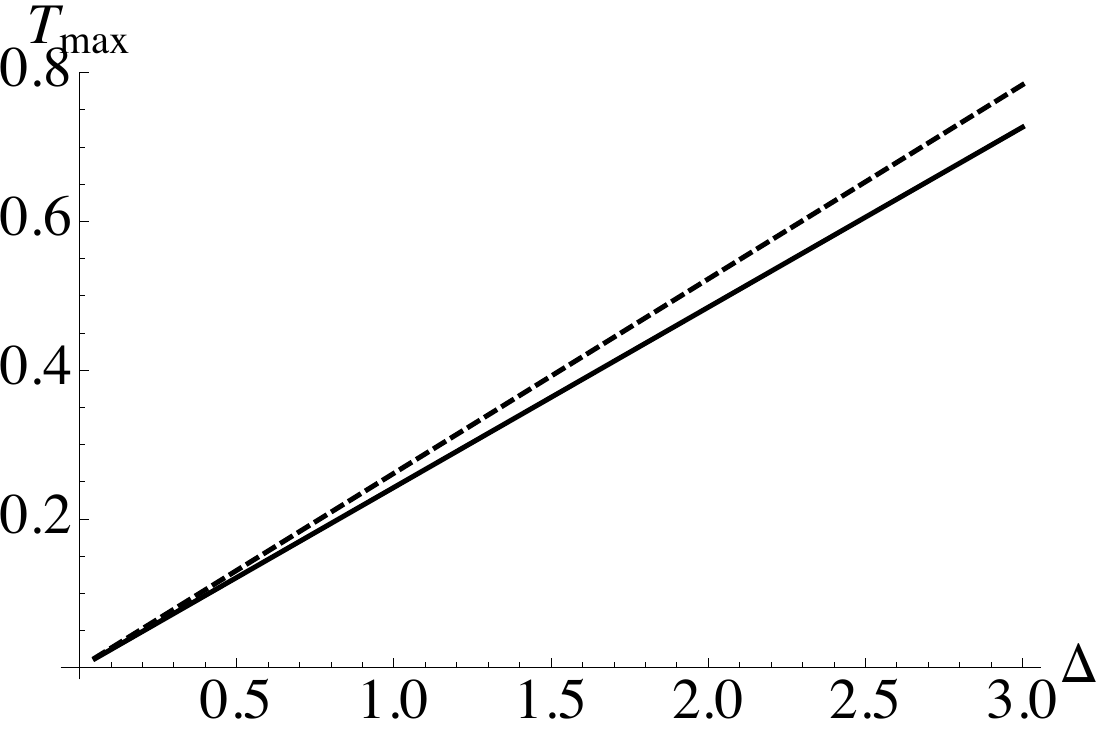}
\caption{{\bf (a)} The QFI for three values of energy spacing, $\Delta$, against temperature, $T$. {\bf (b)} One over the maximum value of QFI, $1/\mathcal{H}_\text{max}$, against $\Delta$ showing the quadratic relationship. {\bf (c)} The linear relationship between the value of temperature, $T_\text{max}$, when $\mathcal{H}_\text{max}$ is achieved plotted against $\Delta$. In all panels solid curves are for the qubit and dashed curves are for the oscillator.}
\label{qfi}
\end{figure}

Remarkably for a family of Gibbs states $\varrho_T$, the QFI can be easily evaluated and is equal to the classical Fisher information corresponding to a measurement described by the eigenstates of the Hamiltonian $H$. In formula one obtains \cite{ZanardiPRA2008,LuisPRL,ParisJPA2016,DePasqualeNatComm2016}
\begin{equation}
\label{eq:QFI}
\rev{\mathcal{H} (T)= \sum_{n=1}^d \frac{|\partial_T p_n |^2}{p_n} = \frac{[\text{Var}(H)]^2}{T^4}\,.}
\end{equation}
For a thermal two level system with free Hamiltonian $H\!=\!\tfrac{\Delta}{2} \sigma_z$ ($\sigma_z$ being the Pauli matrix), the corresponding Gibbs state is
\begin{equation}
\label{eq:rho}
\varrho_s(T) = \frac{1}{2} \left(
\begin{array}{cc}
1-\tanh \left(\frac{\Delta }{2 T}\right)\ & 0 \\
 0 & 1+\tanh \left(\frac{\Delta }{2 T}\right) \\
\end{array}
\right).
\end{equation}
We can determine the QFI using Eq.~\eqref{eq:QFI} and find
\begin{equation}
\label{QFIqubit}
\mathcal{H}(T) = \frac{\Delta ^2 \text{sech}^2\left(\frac{\Delta }{2 T}\right)}{2 T^4}.
\end{equation}
In Fig.~\ref{qfi} {\bf (a)} the solid curves show the QFI for several values of the energy level splitting, $\Delta$. \rev{Clearly, smaller energy gaps can lead to significantly better precision, however an important point to note is that the QFI peaks at a single value of $T$. This means there is a single temperature that a given two-level system with a specified energy gap is optimized to probe. This temperature corresponds to the value of $T$ maximizing the QFI, $\mathcal{H}_\text{max}$, and as we change $\Delta$ the position of this peak shifts.} We find there exists a simple quadratic relation between the ultimate precision\rev{, i.e. $\mathcal{H}_\text{max}$,} and the spacing
\begin{equation}
\frac{1}{\mathcal{H}_\text{max}} = \frac{\sqrt{\pi}}{8} \Delta^2,
\end{equation}
as shown in Fig.~\ref{qfi} {\bf (b)}. Furthermore, the value of temperature corresponding to this maximum, $T_\text{max}$, is linearly related to $\Delta$
\begin{equation}
\label{eq:qubitTmax}
T_\text{max} = \alpha \Delta~\qquad~\text{where}~\qquad~2\alpha = \text{tanh}\left(\frac{1}{\alpha}\right).
\end{equation}
as shown in Fig.~\ref{qfi} {\bf (c)}.
From these relations we see $1/\mathcal{H} \sim \Delta^2 \sim T_{max}^2$ in line with the Landau bound~\cite{ParisJPA2016}.

Moving to the other well studied scenario, we consider the infinite-dimensional quantum harmonic oscillator, $H\!=\!\Delta(a^\dagger a + \tfrac{1}{2})$ ($a$ representing the bosonic annihilation operator satisfying $[a,a^\dagger] = \mathbbm{1}$), with spectral gap $\Delta$. The QFI for a thermal state is given by
\begin{equation}
\label{QFIHO}
\mathcal{H}(T)= \frac{\Delta ^2 \text{csch}^2\left(\frac{\Delta}{2 T}\right)}{4 T^4}.
\end{equation}
The dashed lines in Fig.~\ref{qfi} show that all the features exhibited by the simple two-level system carry over almost identically to this case. In particular, smaller $\Delta$ leads to increased sensitivities at lower temperatures, there is a quadratic relationship between the maximum QFI and $\Delta$
\begin{equation}
\frac{1}{\mathcal{H}_\text{max}} \simeq \frac{\sqrt{\pi}}{5\sqrt{3}} \Delta^2,
\end{equation}
and we find the value of $T$ maximizing the QFI scales linearly with $\Delta$. Note that for both scenarios this linear relation is not surprising since both, the density operator and the resulting QFI, only depend on the ratio $\Delta/T$, but not on $\Delta$ and $T$ separately.

Clearly the two disparate dimensional systems exhibit qualitatively identical behaviors, thus implying that the achievable precision for thermometry with harmonic systems is solely dependent on the single characteristic spectral gap, $\Delta$, while dimensionality plays only a minor role. \rev{In fact, we can show this more explicitly by considering arbitrary $d-$dimensional harmonic systems, described by the Hamiltonian $H\!\!=\!\! \sum_{n=1}^d n\Delta \ket{E_n}\bra{E_n}$, and calculating the corresponding QFI.} For a thermal state, the probe is in Gibbs form and the energy level occupations (eigenvalues) are simply given by a Boltzmann distribution. Thus, for a $d$-dimensional system with energy spacing $\Delta$, the $n$-th eigenvalue \rev{of the thermal state} is
\begin{equation}
\rev{p_n = 
\frac{e^{-\frac{n \Delta}{T}}}{\mathcal{Z}} = \frac{\left(e^{\frac{\Delta }{T}}-1\right) e^{\frac{\Delta(d-n)}{T}}}{e^{\frac{\Delta d}{T}}-1}.}
\end{equation}
From Eq.~\eqref{eq:QFI} we know that the QFI is based solely on the rate of change of these occupations with respect to temperature, and we obtain
\begin{equation}
\mathcal{H}_d(T) = \tfrac{\Delta ^2 \left(d^2 \left(-e^{\frac{\Delta  d}{T}}\right)-d^2 e^{\frac{\Delta  (d+2)}{T}}+2 \left(d^2-1\right) e^{\frac{\Delta (d+1)}{T}}+e^{\frac{\Delta +2 \Delta d}{T}}+e^{\frac{\Delta }{T}}\right)}{T^4 \left(e^{\frac{\Delta}{T}}-1\right)^2 \left(e^{\frac{\Delta d}{T}}-1\right)^2}.
\end{equation}
It is easy to check that we recover Eq.~\eqref{QFIqubit} (Eq.~\eqref{QFIHO}) by setting $d=2$ ($d\to\infty$).

\begin{figure}[t]
\includegraphics[width=0.8\columnwidth]{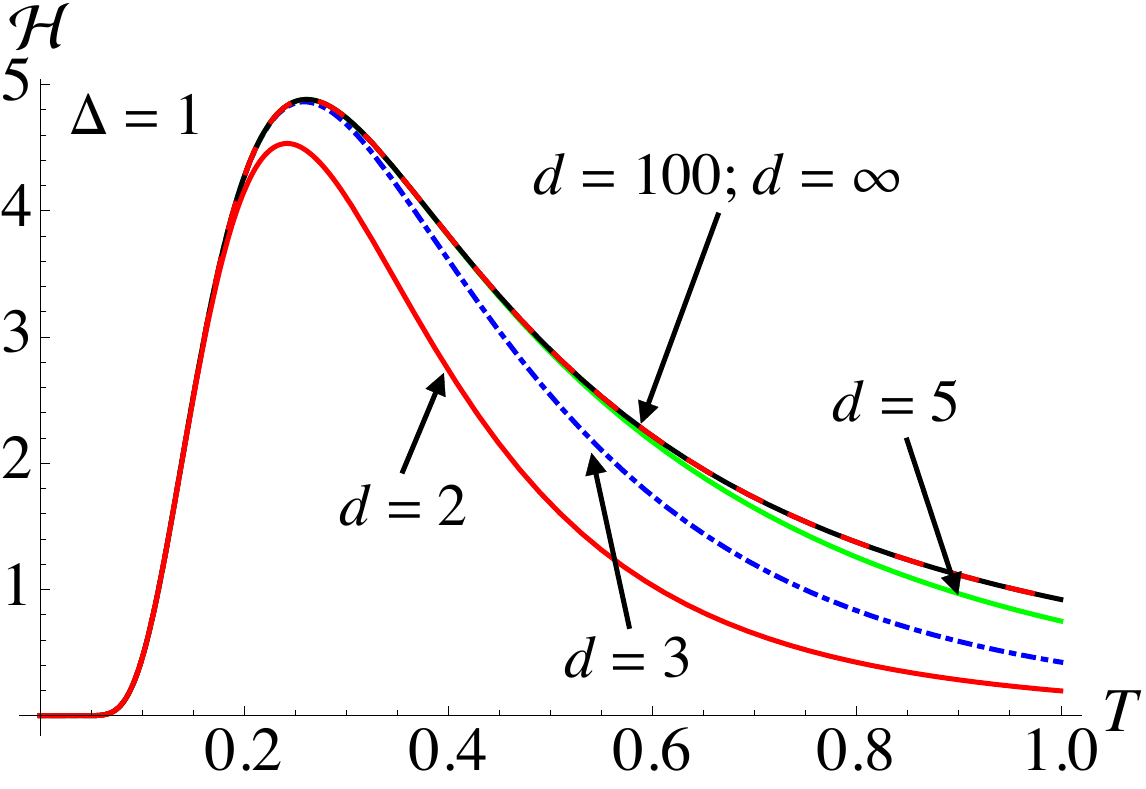}
\caption{QFI for several different dimensional systems with harmonic spectra. We have fixed $\Delta=1.0$.}
\label{arbdimplot}
\end{figure}

We depict the QFI for various values of $d$ in Fig.~\ref{arbdimplot} where we have (arbitrarily) fixed $\Delta=1$. It is immediately evident that for low temperatures, $T\lesssim 0.2$, all systems perform identically, while differences arise only at comparatively large temperatures. Indeed, such a behavior is intuitive: at low temperatures all systems are constrained to the low energy portion of the spectrum, thus in this region only the ground and first excited state will play a significant role. Of course, the general behavior shown previously, namely quadratic relation between the inverse of $\mathcal{H}_\text{max}$ and $\Delta$ and the linear relationship between $T_\text{max}$ and $\Delta$, persist here. However, it is interesting to notice that qualitatively nothing changes for $d\geq3$ with regards to the maximal precision. 

At this point we can conclude (i) the constant energy level spacing in harmonically gapped systems plays the most crucial role in thermometry. Therefore, to probe low temperatures one should seek to use a system with a small energy spacing, while for larger temperatures larger gapped systems are significantly more useful. (ii) We can gain some enhancement by going from two- to a three-level system, however higher dimensional systems offer no advantage regarding the optimal achievable precision. (iii) Regardless, such systems are only designed to estimate a single temperature with the optimal precision.

Thus, it is then interesting to ask under what conditions more than a one temperature can be accurately probed using a single system. Clearly, when the probe exhibits a unique characteristic energy gap such a situation is impossible. Therefore, we continue with a different setting and consider thermometry using systems with arbitrary energy level spacings.

\section{Thermometry in Arbitrarily Gapped Spectra} \label{s:anharmonic}

The harmonic oscillator is certainly the best studied quantum system, since the dynamics are analytically solvable \cite{Husimi1953} and many potentials can be well-approximated by a harmonic potential for small excitations. \rev{Nevertheless, real systems are rarely exactly described by harmonic oscillators and therefore, due to the additional characteristic energy spacings present in nonlinear systems, one might expect the best precision in thermometry to be intimately related to these features.} While in principle we could individually analyse a given nonlinear system explicitly, it is in fact sufficient to simply consider a system with arbitrary spaced energy levels, where without loss of generality we fix the ground state energy to be zero, since thermometry is only concerned with an eigenstates rate of change rather than the actual eigenenergy value. Evidently, analysing such systems is only meaningful for $d\geq3$ and therefore to begin we will restrict ourselves to $d=3$. 

\begin{figure}[t]
\includegraphics[width=0.8\columnwidth]{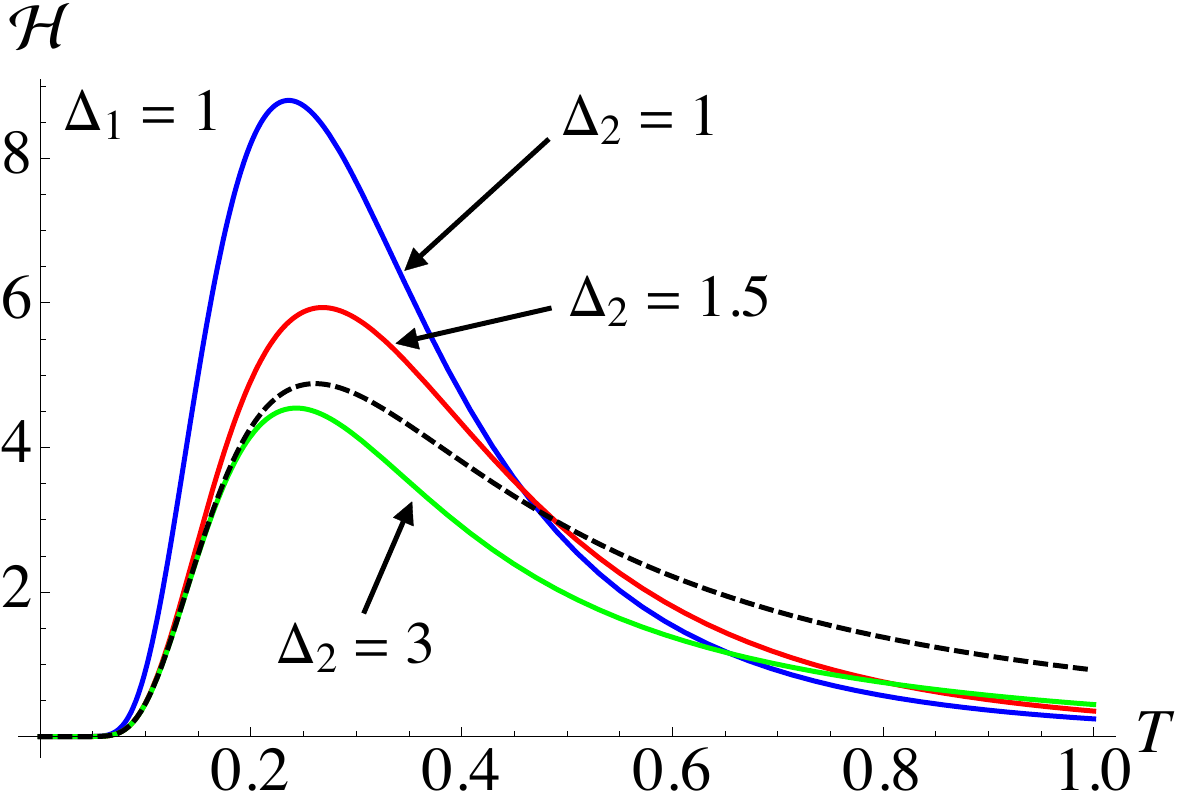}
\caption{QFI for temperature taking several values of energy level splittings in an arbitrarily gapped three level system, Eq.~\eqref{3levelarb}. \rev{For comparison, the black dashed curve is the QFI for the harmonic oscillator, corresponding to the largest QFI achievable for harmonic spectra [cfr. Fig.~\ref{arbdimplot}]}. }
\label{fig:3level}
\end{figure}
For the considered three level system the Hamiltonian is
\begin{equation}
\label{3levelarb}
H=\begin{pmatrix}
 \Delta_2 & 0 & 0 \\
 0 & \Delta_1 & 0 \\
 0 & 0 & 0 \\
\end{pmatrix},
\end{equation}
where the energy gap between the ground (first) and the first (second) excited state is $\Delta_1$ ( $\Delta_2$), and we assume $\Delta_2 \geq \Delta_1$. We can readily evaluate the Gibbs state and therefore the corresponding QFI. In Fig.~\ref{fig:3level} we show the QFI when we fix $\Delta_1=1$ and  take several values of $\Delta_2$. The topmost blue curve for $\Delta_2=1$ corresponds to a doubly degenerate excited state, showing that a significantly better precision is achievable and the temperature which it most accurately estimates is exactly the same as for the qubit case found by solving Eq.~\eqref{eq:qubitTmax}. This is precisely the result rigorously proven by Correa \etal~\cite{LuisPRL}, wherein it was shown that the optimal probe corresponds to a system with a highly degenerate excited state, and the precision is enhanced as one includes more degenerate energy levels. It should be noted however, that again this system can only optimally estimate a single temperature  due to the single characteristic energy spacing. By increasing $\Delta_2$ we see that its main effect is to reduce the optimal precision and to introduce a slight shift in the position of $\mathcal{H}_\text{max}$. For $\Delta_2=1.5$, hence the gap is less than that of a three level harmonic spectrum, we see the precision is still larger than the oscillator case for $T<0.5$ and clearly this situation interpolates between the optimal degenerate probe ($\Delta_2=1$) and the harmonic case ($\Delta_2=2$). Taking $\Delta_2=3$, we find that for low temperatures the behavior is indistinguishable from the two-level case; this due to the fact that, at such values of $T$, thermal energy is not sufficient to excite the system up to the second excited state, and one has essentially a two-level system. The effect of the second excited state only becomes apparent for large $T$, where one eventually obtains values of the QFI larger than the other three-level systems, however never outperforming the harmonic oscillator case. 

Clearly the gap between ground state and first excited state still plays the most dominant role, and furthermore, even for arbitrary spacings, all these systems still exhibit a single maxima which is primarily dictated by $\Delta_1$. Therefore, even including more energy levels with arbitrary gaps between them, we achieve the same behavior as shown in Fig.~\ref{fig:3level}, leading to the conclusion that {\it any single} quantum system with {\it non-degenerate} energy spectrum can only accurately determine a single temperature.

\subsection{Introducing Degeneracy}
The increased sensitivity achieved by employing degenerate energy levels in Ref.~\cite{LuisPRL} relies on the fact that for a system with such a single energy gap, all the degenerate levels begin to become populated simultaneously and thus the witnessed enhancement. Naturally then, we ask if a similar approach can be used to allow a single system to probe more than one temperature, and if so how much degeneracy is required.

\begin{figure}[t]
{\bf (a)}\\ 
\includegraphics[width=0.8\columnwidth]{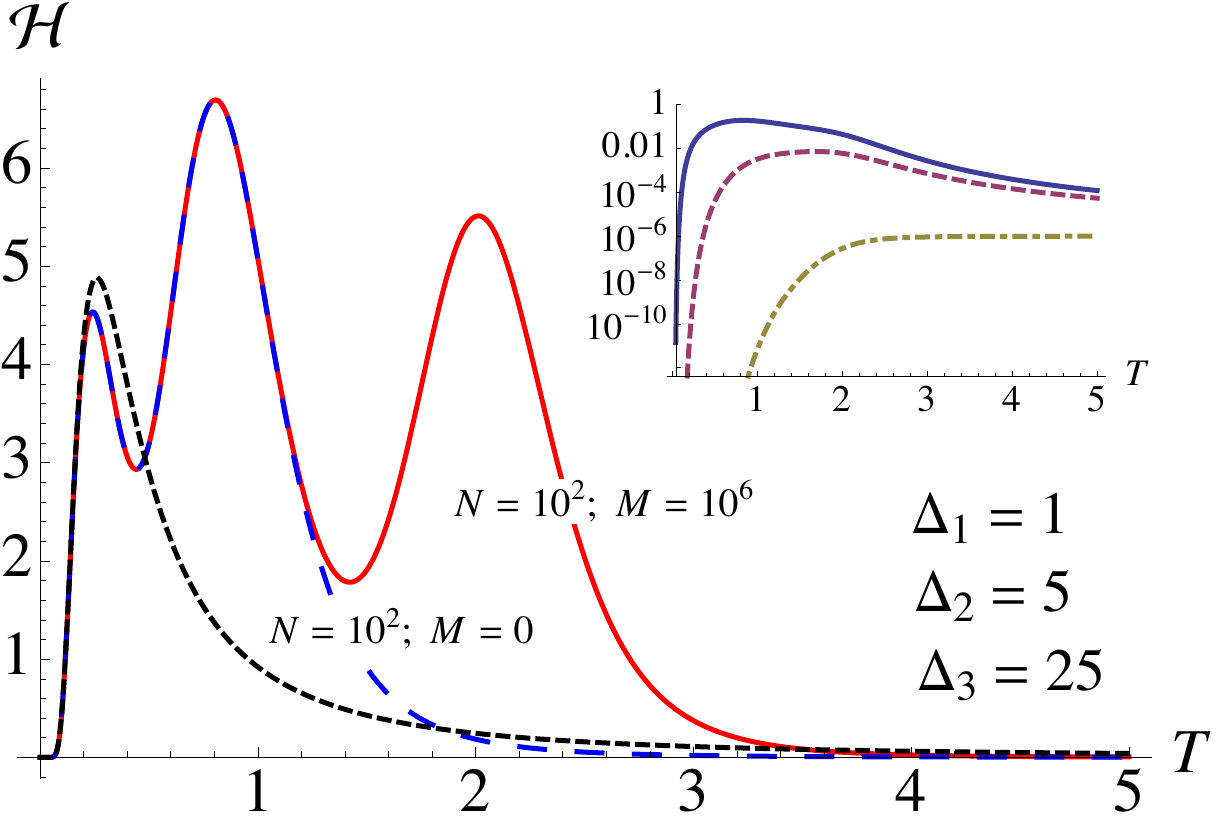}\\
{\bf (b)}\\
\includegraphics[width=0.8\columnwidth]{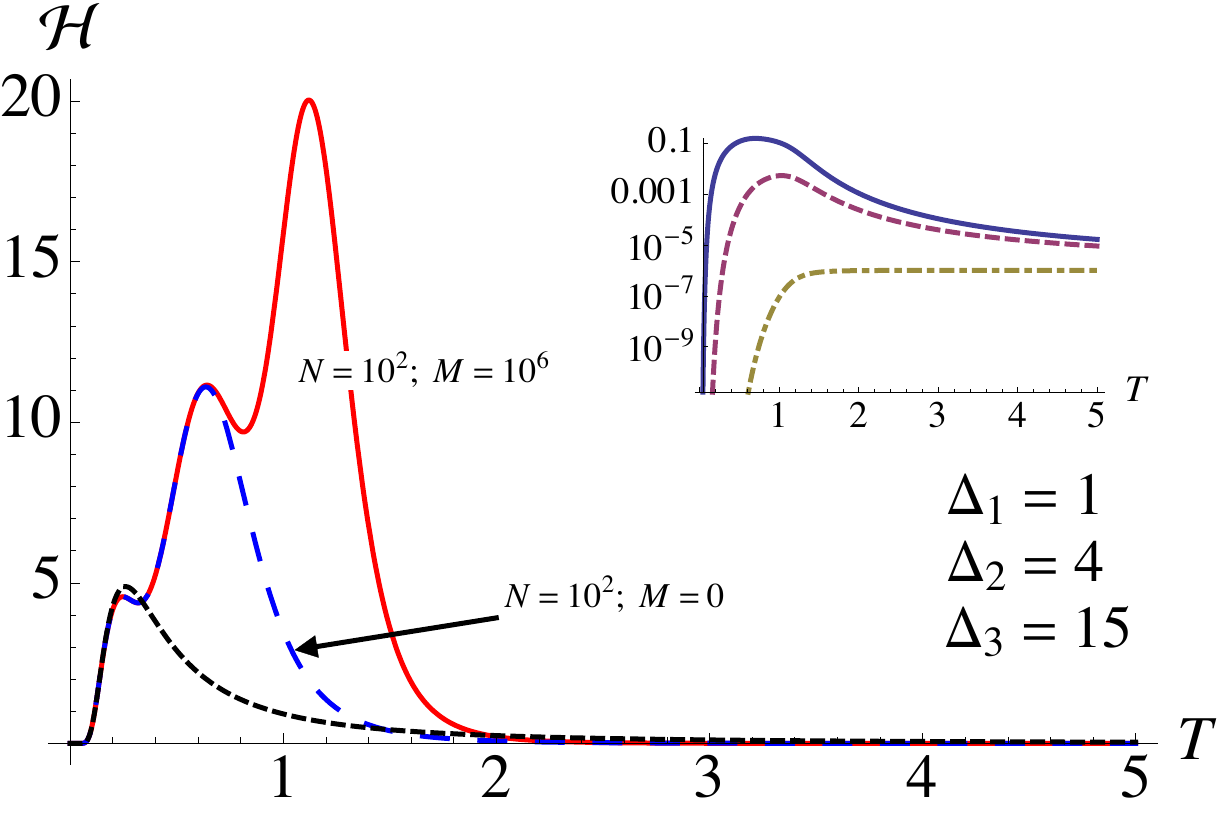}
\caption{\rev{QFI for the Gibbs state of systems with highly degenerate anharmonic spectra given in Eq.~\eqref{spectra}.} {\bf (a)} With large gaps, $\Delta_1=1$, $\Delta_2=5$, and $\Delta_3=25$. {\bf (b)} With smaller gaps $\Delta_1=1$, $\Delta_2=4$, and $\Delta_3=15$. In both panels the black dashed curve corresponds to the QFI for the harmonic oscillator for comparison. {\it Insets:} The population of the first exited state [topmost, solid], and one of the second [middle, dashed] and third [bottommost, dot-dashed] excited states.}
\label{fig:gaps}
\end{figure}

Consider a system with the following energy spectrum
\rev{\begin{eqnarray}
\label{spectra}
&E_0 &=0 \nonumber \\
&E_1 &= \Delta_1 \nonumber \\
&E_i &= \Delta_2~~\text{for}~~N+2\geq i \geq2 \\
&E_j &= \Delta_3~~\text{for}~~ M+N+2 \geq j\geq n+3\nonumber  \\
& &\vdots \nonumber 
\end{eqnarray}}
Here a unique ground and first excited state are separated by an energy splitting $\Delta_1$. A gap of $\Delta_2$ separates the first excited state from an $N$-fold degenerate second excited state, which is in turn separated from an $M$-fold degenerate third excited state by $\Delta_3$. In Fig.~\ref{fig:gaps} we fix $\Delta_1\!=\!1$ and study the magnitude of the gap and amount of degeneracy needed to resolve more temperatures. By taking a suitably large spacing between the $\Delta$'s, the QFI now exhibits more peaks (see panel {\bf (a)}), however this comes at the requirement for high degeneracy. While the first peak is almost indistinguishable from the two-level system case, we require two-orders of magnitude more degenerate second excited states to resolve a second temperature with a comparable accuracy, and of the order of $10^6$ states if we wish to resolve a third. From panel {\bf (b)} we see that this behavior is delicately dependent on the size of the respective gaps. Taking splittings closer together we find that the system tends to be more sensitive at higher temperatures and the range of temperatures it can reasonably accurately determine, when compared to the magnitude of the first peak, is significantly enlarged. We see the reason for this by examining the insets of Fig.~\ref{fig:gaps}. Larger splittings means the highly degenerate excited states exhibit significant rates of change only when the ground and first excited state are by comparison stable in population. Contrarily, for smaller splittings we find all energy levels exhibit more marked rates of change at overlapping temperatures. Thus the high degree of degeneracy enhances these effects at such temperatures.

\section{Thermometry and the Quantum Speed Limit} \label{s:QSL}
The preceding sections crucially assumed that the probe system measured had already thermalized to a canonical Gibbs state. Here we relax this assumption and explore a complementary question: are there any signatures of the optimal temperature a given probe is able to estimate present in the dynamics of a thermometry protocol? In a similar context, it has been shown that the quantum speed limit determines the best possible precision in quantum metrology \cite{LloydPRL2006,LloydPRL2012}. For thermal states, all occupation probabilities are simply functions of the temperature. Thus, it is not far-fetched to assume that some kind of resonance between the eigen-dynamics of the probe and the temperature of the environment (i.e. the probed system) determines a ``preferred'' time-scale of the joint dynamics.

To address this, we use the quantum speed limit (QSL) to quantitatively explore the dynamical features of a simple thermometry scheme where the probe system, $\varrho_s$, is placed in contact with the thermal environment, $\varrho_E$. Note that it is a well established fact that the QSL can be expressed in terms of the QFI corresponding to time $t$ estimation~\cite{TaddeiPRL}. Therefore while a clear, albeit somewhat trivial, connection could be made, this does not necessarily imply a proper relation of the QSL with quantum thermometry. To get clearer insight into the matter we now compute the QSL for thermalization processes, and analyze whether extrema in the quantum speed correspond to maxima of QFI for temperature.

We will focus primarily on the formulation of the QSL provided in Ref.~\cite{DeffnerPRL}, where the speed for an arbitrary process is bounded by
\begin{equation}
\label{vQSL}
v \leq v_\text{QSL}=\frac{\| \mathcal{D}(\varrho_s) \|_\text{op}}{\co{\mathcal{B}} \si{\mathcal{B}}},
\end{equation}
where $\mathcal{D}(\cdot)$ is the generator of the dynamics, $\| \cdot \|_\text{op}$ is the operator norm, and $\mathcal{B}=\arccos(F(\rho_0, \rho_t))$ is the Bures angle between the initial and time-evolved states~\cite{DeffnerPRL}. From this maximal quantum speed we obtain the corresponding QSL time \cite{DeffnerPRL},
\begin{equation}
\label{tQSL}
\tau_\text{QSL}=\frac{\sin^2(\mathcal{B})}{2 E_\tau},
\end{equation}
where $E_\tau$ is the time-averaged norm of the generator, $E_\tau=1/\tau\,\int_0^\tau\, dt\, \| \mathcal{D}(\varrho_s(t)) \|_\text{op}$ and $\tau$ is the total evolution time-window. 

As established in Ref.~\cite{LuisPRL} the optimal initial configuration for the probe system is in its ground state. In the following, we consider this initial condition and study two types of environment: (i) an infinite reservoir within the Markovian limit such that the probe system thermalizes to the Gibbs state, and (ii) a finite dimensional environment such that the probe system periodically reaches the Gibbs state. We restrict our probe to be a two-level system, however qualitatively the results are unaffected for $d>2$.

\begin{figure}[t]
{\bf (a)}\\ 
\includegraphics[width=0.8\columnwidth]{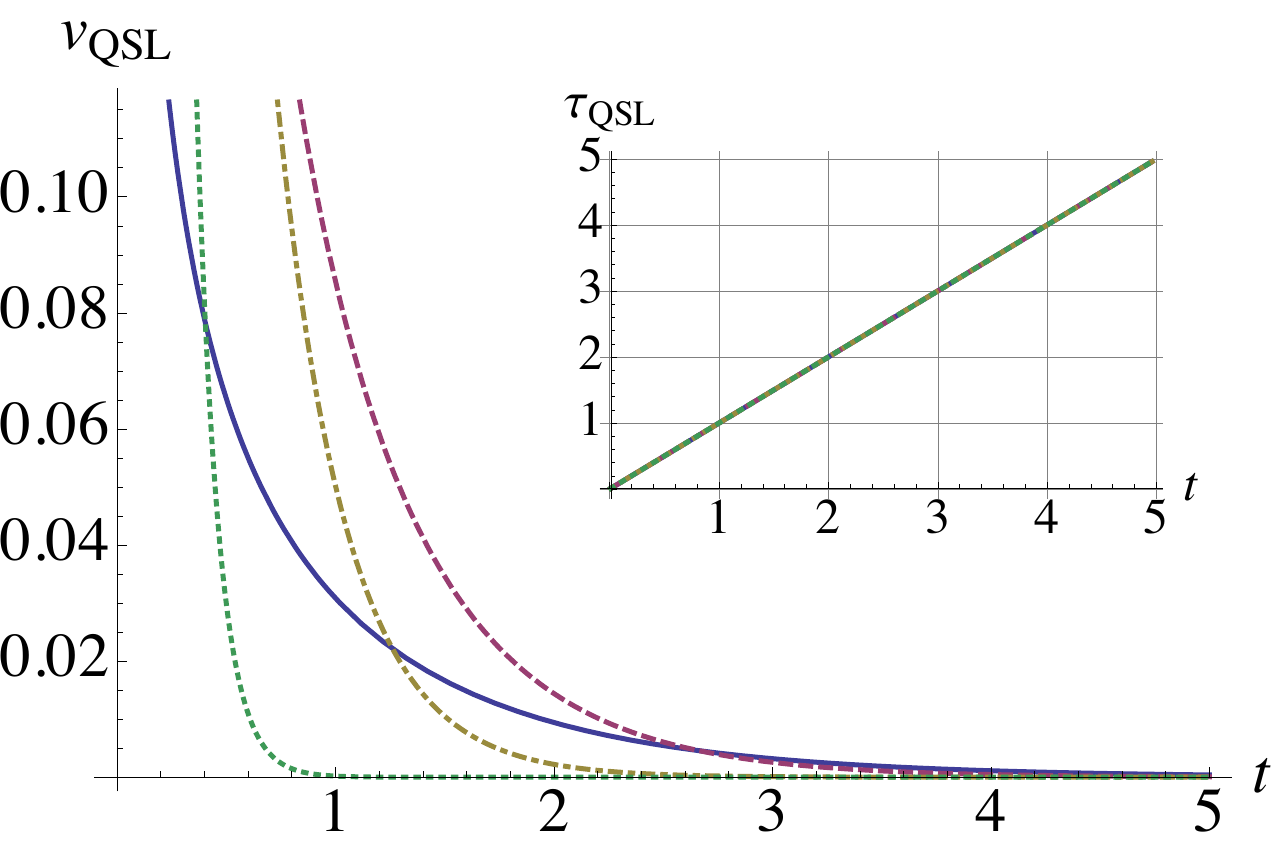}\\
{\bf (b)}\\
\includegraphics[width=0.8\columnwidth]{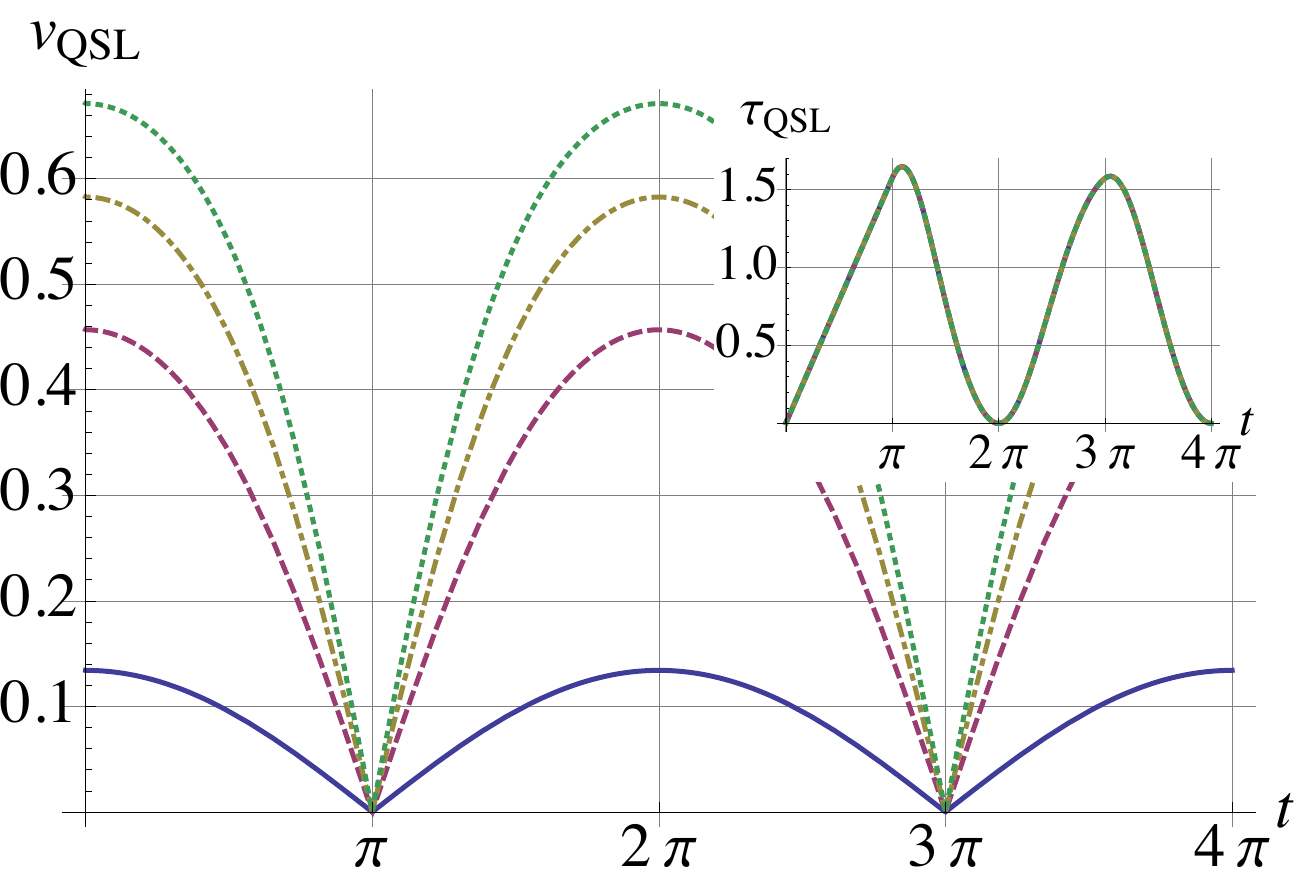}
\caption{{\bf (a)} QSL for an infinite dimensional thermal environment, fixing $\Delta\!=\!\gamma\!=\!1$ with $T=0.25$ (solid), 0.75 (dashed), 1.5 (dot-dashed), 5 (dotted). {\bf (b)} QSL against time for a finite dimensional environment fixing $J\!=\!\Delta\!=\!1$ and taking $T=0.25$, 0.75, 1.5, 5 (bottom to top curves). In both panels the insets show the respective QSL times, $\tau_\text{QSL}$, \rev{Eq.~\eqref{tQSL}, where all curves are identical regardless of $T$.}}
\label{fig5}
\end{figure}
\subsection{Infinite Dimensional Environment}
A two-level probe coupled to a infinite thermal bath can be effectively modelled using a master equation in Lindblad form
\begin{equation}
\label{eq:master}
\dot{\varrho}_s(t) = \mathcal{D}(\varrho_s(t)) = -i\left[ H_s , \varrho_s(t) \right] + \mathcal{L}(\varrho_s(t)),
\end{equation}
with 
\begin{equation}
\begin{aligned}
\mathcal{L}(\rho) = & \gamma (\nbar + 1) \Big[\sigma_- \rho \sigma_+ - \{\rho, \sigma_+\sigma_- \} \Big] \\
			    &+ \gamma \nbar \Big[\sigma_+ \rho \sigma_- - \{ \rho, \sigma_- \sigma_+ \} \Big].
\end{aligned}
\end{equation}
and where \rev{$\nbar=1/(e^{\frac{\Delta }{T}}-1)$}. Setting $\varrho_s(0)=\ket{0}\bra{0}$ we can readily solve Eq.~\eqref{eq:master} and therefore evaluate Eqs.~\eqref{vQSL} and \eqref{tQSL}. In fact, in this case we find the speed matches exactly with the one derived by extending the Mandelstam-Tamm bound based on an approach involving the QFI corresponding to the estimation of the evolution time $t$ of the probe state \cite{TaddeiPRL}. Therefore in this setting the bound is tight since the two approaches coincide, and we find
\begin{equation}
v_\text{QSL} = \frac{\gamma t(1+2\nbar) e^{-(1+2\nbar)}}{2 \sqrt{\nbar(1+\nbar - e^{-2(1+2\nbar)\gamma t})(e^{-2(1+2\nbar)\gamma t})+\nbar)}}.
\end{equation}
In Fig.~\ref{fig5} {\bf (a)} we show $v_\text{QSL}$ for various values of bath temperature, $T$. Clearly, when the system is thermalized $v_\text{QSL}\to 0$. However, this is to be expected since the thermal state is a fixed point of the dynamical map. We see the speed tends to reduce at a faster rate for larger values of $T$, however this effect plays no role in the ability to precisely estimate the temperature of the environment. In fact, this is further evidenced in calculating the QSL time shown in the inset. Here we find regardless of the parameters chosen the QSL time is unaffected. This would appear at variance with the intuition posited previously, \rev{i.e. that the dynamics would reflect the optimal temperature a system was `designed' to probe.} While in the figure we have shown $\Delta\!=\!\gamma\!=\!1$, we remark that qualitatively the same behavior is exhibited for all parameter choices.

\subsection{Finite Dimensional Environment}
Changing the environment to be finite dimensional will introduce a degree of non-Markovianity to the dynamics. In what follows we consider the case when the dimensionality of both probe and environment are equal, i.e. $d_s\!=\!d_E\!=\!2$. In this situation we must carefully define the interaction to ensure that the probe reaches the thermal state at some point during the dynamics. This can be realized by using an excitation preserving interaction and the free evolution terms
\begin{equation}
\label{eq:thermo}
H_s= H_E = \frac{\Delta}{2} \sigma^{s(E)}_z,~\qquad~H_{int} = \frac{J}{4} (\sigma^s_x \otimes \sigma^E_x + \sigma^s_y \otimes \sigma^E_y),
\end{equation}
such that at $t=\pi/J$ the thermal state of the environment is swapped with the state of the probe. Equation~\eqref{eq:thermo} can be understood as a simple version of a thermometer, that is similar in size to the quantum system of interest.

In order to evaluate Eqs.~\eqref{vQSL} and \eqref{tQSL}, we require the generator of the dynamics. While in principle one could derive the associated equations of motion, given the comparative simplicity of the setting we can readily determine $\varrho_s(t)$ directly, and therefore $\dot{\varrho}_s(t)$. In Fig.~\ref{fig5} {\bf (b)} we show $v_\text{QSL}$ for the same values of the environmental temperature as in panel {\bf (a)}. Immediately, several similarities arise. In particular, we observe that when the probe is in the Gibbs state the speed $v_\text{QSL} = 0$. Furthermore, while different values of the environmental temperature exhibit different dynamical speeds, again there is no indication of a ``preferred" temperature. The inset shows, inline with the infinite dimensional environment case, that the associated $\tau_\text{QSL}$ is identical for all values of $T$. Again, qualitatively identical results hold for any choice of parameters. 

\section{Concluding remarks} \label{s:conclusions}
We have discussed precision thermometry via individual quantum systems. First we have focused on quantum probes that have already thermalized, evaluating the ultimate precision achievable in temperature estimation for quantum systems characterized by diverse energy spectra. Starting with equally gapped spectra described by a single energy spacing $\Delta$, we have demonstrated that for any dimension of the quantum system there is a single optimal temperature scaling linearly with $\Delta$, \rev{confirming and extending to arbitrary dimension, the results obtained previously in \cite{LuisPRL,ParisJPA2016}}. We have then shown that only by introducing anharmonicity to a system with high dimensionality and high degeneracy, is it possible to \rev{go beyond these results and} estimate more than one temperature with high accuracy, i.e. one obtains a QFI with more than single peak. Additionally in this case we have discussed in detail the relationship between the energy splittings $\Delta_j$ and the temperatures that can be estimated efficiently. In particular we observed that taking splittings close together greatly enhances the estimation of higher temperatures.

Given that at equilibrium the temperature determines the average energy of a system, we have analyzed the relationship between thermometry and the QSL. Whereas for single systems in equilibrium the temperature and the QSL are trivially related (since both are essentially estimated by the average energy), we falsified the hypothesis that the QSL sets a preferred precision in dynamical measurements. To this end, we studied paradigmatic examples: systems relaxing with an infinite heat reservoir and a finite thermometer. While the QSL fully characterizes the thermalization dynamics, no clear relation to an extremum of the QFI for temperature was unveiled.

\acknowledgements{ MGG acknowledges support from Marie Sk\l odowska-Curie
Action H2020-MSCA-IF-2015 (project ConAQuMe, grant no. 701154). SD acknowledges support from the U.S. National Science Foundation under Grant No. CHE-1648973.}

\bibliography{ThermometryQSL}

\begin{thebibliography}{30}%
\makeatletter
\providecommand \@ifxundefined [1]{%
 \@ifx{#1\undefined}
}%
\providecommand \@ifnum [1]{%
 \ifnum #1\expandafter \@firstoftwo
 \else \expandafter \@secondoftwo
 \fi
}%
\providecommand \@ifx [1]{%
 \ifx #1\expandafter \@firstoftwo
 \else \expandafter \@secondoftwo
 \fi
}%
\providecommand \natexlab [1]{#1}%
\providecommand \enquote  [1]{``#1''}%
\providecommand \bibnamefont  [1]{#1}%
\providecommand \bibfnamefont [1]{#1}%
\providecommand \citenamefont [1]{#1}%
\providecommand \href@noop [0]{\@secondoftwo}%
\providecommand \href [0]{\begingroup \@sanitize@url \@href}%
\providecommand \@href[1]{\@@startlink{#1}\@@href}%
\providecommand \@@href[1]{\endgroup#1\@@endlink}%
\providecommand \@sanitize@url [0]{\catcode `\\12\catcode `\$12\catcode
  `\&12\catcode `\#12\catcode `\^12\catcode `\_12\catcode `\%12\relax}%
\providecommand \@@startlink[1]{}%
\providecommand \@@endlink[0]{}%
\providecommand \url  [0]{\begingroup\@sanitize@url \@url }%
\providecommand \@url [1]{\endgroup\@href {#1}{\urlprefix }}%
\providecommand \urlprefix  [0]{URL }%
\providecommand \Eprint [0]{\href }%
\providecommand \doibase [0]{http://dx.doi.org/}%
\providecommand \selectlanguage [0]{\@gobble}%
\providecommand \bibinfo  [0]{\@secondoftwo}%
\providecommand \bibfield  [0]{\@secondoftwo}%
\providecommand \translation [1]{[#1]}%
\providecommand \BibitemOpen [0]{}%
\providecommand \bibitemStop [0]{}%
\providecommand \bibitemNoStop [0]{.\EOS\space}%
\providecommand \EOS [0]{\spacefactor3000\relax}%
\providecommand \BibitemShut  [1]{\csname bibitem#1\endcsname}%
\let\auto@bib@innerbib\@empty
\bibitem [{\citenamefont {{Callen}}(1985)}]{Callen1985}%
  \BibitemOpen
  \bibfield  {author} {\bibinfo {author} {\bibfnamefont {H.~B.}\ \bibnamefont
  {{Callen}}},\ }\href@noop {} {\emph {\bibinfo {title} {{Thermodynamics and an
  Introduction to Thermostatistics, 2nd Edition}}}}\ (\bibinfo  {publisher}
  {Wiley},\ \bibinfo {year} {1985})\BibitemShut {NoStop}%
\bibitem [{\citenamefont {Gemmer}\ \emph {et~al.}(2009)\citenamefont {Gemmer},
  \citenamefont {Michel},\ and\ \citenamefont {Mahler}}]{Gemmer2009}%
  \BibitemOpen
  \bibfield  {author} {\bibinfo {author} {\bibfnamefont {Jochen}\ \bibnamefont
  {Gemmer}}, \bibinfo {author} {\bibfnamefont {M.}~\bibnamefont {Michel}}, \
  and\ \bibinfo {author} {\bibfnamefont {G\"{u}nter}\ \bibnamefont {Mahler}},\
  }\href@noop {} {\emph {\bibinfo {title} {{Quantum Thermodynamics}}}}\
  (\bibinfo  {publisher} {Springer},\ \bibinfo {address} {Berlin /
  Heidelberg},\ \bibinfo {year} {2009})\BibitemShut {NoStop}%
\bibitem [{\citenamefont {Helstrom}(1976)}]{HelstromBook}%
  \BibitemOpen
  \bibfield  {author} {\bibinfo {author} {\bibfnamefont {Carl~W.}\ \bibnamefont
  {Helstrom}},\ }\href@noop {} {\emph {\bibinfo {title} {{Quantum Detection and
  Estimation Theory}}}}\ (\bibinfo  {publisher} {Academic Press},\ \bibinfo
  {address} {New York},\ \bibinfo {year} {1976})\BibitemShut {NoStop}%
\bibitem [{\citenamefont {Paris}(2009)}]{ParisIJQI2009}%
  \BibitemOpen
  \bibfield  {author} {\bibinfo {author} {\bibfnamefont {M.~G.~A.}\
  \bibnamefont {Paris}},\ }\bibfield  {title} {\enquote {\bibinfo {title}
  {Quantum estimation for quantum technology},}\ }\href
  {http://www.worldscientific.com/doi/abs/10.1142/S0219749909004839} {\bibfield
   {journal} {\bibinfo  {journal} {Int. J. Quant. Inf.}\ }\textbf {\bibinfo
  {volume} {07}},\ \bibinfo {pages} {125} (\bibinfo {year} {2009})}\BibitemShut
  {NoStop}%
\bibitem [{\citenamefont {Stace}(2010)}]{Stace2010}%
  \BibitemOpen
  \bibfield  {author} {\bibinfo {author} {\bibfnamefont {T.~M.}\ \bibnamefont
  {Stace}},\ }\bibfield  {title} {\enquote {\bibinfo {title} {Quantum limits of
  thermometry},}\ }\href {\doibase 10.1103/PhysRevA.82.011611} {\bibfield
  {journal} {\bibinfo  {journal} {Phys. Rev. A}\ }\textbf {\bibinfo {volume}
  {82}},\ \bibinfo {pages} {011611} (\bibinfo {year} {2010})}\BibitemShut
  {NoStop}%
\bibitem [{\citenamefont {Monras}\ and\ \citenamefont
  {Illuminati}(2011)}]{Monras2011}%
  \BibitemOpen
  \bibfield  {author} {\bibinfo {author} {\bibfnamefont {A.}~\bibnamefont
  {Monras}}\ and\ \bibinfo {author} {\bibfnamefont {F.}~\bibnamefont
  {Illuminati}},\ }\bibfield  {title} {\enquote {\bibinfo {title} {Measurement
  of damping and temperature: Precision bounds in gaussian dissipative
  channels},}\ }\href {\doibase 10.1103/PhysRevA.83.012315} {\bibfield
  {journal} {\bibinfo  {journal} {Phys. Rev. A}\ }\textbf {\bibinfo {volume}
  {83}},\ \bibinfo {pages} {012315} (\bibinfo {year} {2011})}\BibitemShut
  {NoStop}%
\bibitem [{\citenamefont {Correa}\ \emph {et~al.}(2015)\citenamefont {Correa},
  \citenamefont {Mehboudi}, \citenamefont {Adesso},\ and\ \citenamefont
  {Sanpera}}]{LuisPRL}%
  \BibitemOpen
  \bibfield  {author} {\bibinfo {author} {\bibfnamefont {L.~A.}\ \bibnamefont
  {Correa}}, \bibinfo {author} {\bibfnamefont {M.}~\bibnamefont {Mehboudi}},
  \bibinfo {author} {\bibfnamefont {G.}~\bibnamefont {Adesso}}, \ and\ \bibinfo
  {author} {\bibfnamefont {A.}~\bibnamefont {Sanpera}},\ }\bibfield  {title}
  {\enquote {\bibinfo {title} {Individual quantum probes for optimal
  thermometry},}\ }\href {\doibase 10.1103/PhysRevLett.114.220405} {\bibfield
  {journal} {\bibinfo  {journal} {Phys. Rev. Lett.}\ }\textbf {\bibinfo
  {volume} {114}},\ \bibinfo {pages} {220405} (\bibinfo {year}
  {2015})}\BibitemShut {NoStop}%
\bibitem [{\citenamefont {Brunelli}\ \emph {et~al.}(2011)\citenamefont
  {Brunelli}, \citenamefont {Olivares},\ and\ \citenamefont
  {Paris}}]{BrunelliPRA2011}%
  \BibitemOpen
  \bibfield  {author} {\bibinfo {author} {\bibfnamefont {M.}~\bibnamefont
  {Brunelli}}, \bibinfo {author} {\bibfnamefont {S.}~\bibnamefont {Olivares}},
  \ and\ \bibinfo {author} {\bibfnamefont {M.~G.~A.}\ \bibnamefont {Paris}},\
  }\bibfield  {title} {\enquote {\bibinfo {title} {Qubit thermometry for
  micromechanical resonators},}\ }\href {\doibase 10.1103/PhysRevA.84.032105}
  {\bibfield  {journal} {\bibinfo  {journal} {Phys. Rev. A}\ }\textbf {\bibinfo
  {volume} {84}},\ \bibinfo {pages} {032105} (\bibinfo {year}
  {2011})}\BibitemShut {NoStop}%
\bibitem [{\citenamefont {Brunelli}\ \emph {et~al.}(2012)\citenamefont
  {Brunelli}, \citenamefont {Olivares}, \citenamefont {Paternostro},\ and\
  \citenamefont {Paris}}]{BrunelliPRA2012}%
  \BibitemOpen
  \bibfield  {author} {\bibinfo {author} {\bibfnamefont {M.}~\bibnamefont
  {Brunelli}}, \bibinfo {author} {\bibfnamefont {S.}~\bibnamefont {Olivares}},
  \bibinfo {author} {\bibfnamefont {M.}~\bibnamefont {Paternostro}}, \ and\
  \bibinfo {author} {\bibfnamefont {M.~G.~A.}\ \bibnamefont {Paris}},\
  }\bibfield  {title} {\enquote {\bibinfo {title} {Qubit-assisted thermometry
  of a quantum harmonic oscillator},}\ }\href {\doibase
  10.1103/PhysRevA.86.012125} {\bibfield  {journal} {\bibinfo  {journal} {Phys.
  Rev. A}\ }\textbf {\bibinfo {volume} {86}},\ \bibinfo {pages} {012125}
  (\bibinfo {year} {2012})}\BibitemShut {NoStop}%
\bibitem [{\citenamefont {Sab{\'\i}n}\ \emph {et~al.}(2014)\citenamefont
  {Sab{\'\i}n}, \citenamefont {White}, \citenamefont {Hackermuller},\ and\
  \citenamefont {Fuentes}}]{Sabin2014}%
  \BibitemOpen
  \bibfield  {author} {\bibinfo {author} {\bibfnamefont {C.}~\bibnamefont
  {Sab{\'\i}n}}, \bibinfo {author} {\bibfnamefont {A.}~\bibnamefont {White}},
  \bibinfo {author} {\bibfnamefont {L.}~\bibnamefont {Hackermuller}}, \ and\
  \bibinfo {author} {\bibfnamefont {I.}~\bibnamefont {Fuentes}},\ }\bibfield
  {title} {\enquote {\bibinfo {title} {Impurities as a quantum thermometer for
  a bose-einstein condensate},}\ }\href {\doibase 10.1038/srep06436} {\bibfield
   {journal} {\bibinfo  {journal} {Sci. Rep.}\ }\textbf {\bibinfo {volume}
  {4}},\ \bibinfo {pages} {6436} (\bibinfo {year} {2014})}\BibitemShut
  {NoStop}%
\bibitem [{\citenamefont {Jevtic}\ \emph {et~al.}(2015)\citenamefont {Jevtic},
  \citenamefont {Newman}, \citenamefont {Rudolph},\ and\ \citenamefont
  {Stace}}]{Jevtic2015}%
  \BibitemOpen
  \bibfield  {author} {\bibinfo {author} {\bibfnamefont {S.}~\bibnamefont
  {Jevtic}}, \bibinfo {author} {\bibfnamefont {D.}~\bibnamefont {Newman}},
  \bibinfo {author} {\bibfnamefont {T.}~\bibnamefont {Rudolph}}, \ and\
  \bibinfo {author} {\bibfnamefont {T.~M.}\ \bibnamefont {Stace}},\ }\bibfield
  {title} {\enquote {\bibinfo {title} {Single-qubit thermometry},}\ }\href
  {\doibase 10.1103/PhysRevA.91.012331} {\bibfield  {journal} {\bibinfo
  {journal} {Phys. Rev. A}\ }\textbf {\bibinfo {volume} {91}},\ \bibinfo
  {pages} {012331} (\bibinfo {year} {2015})}\BibitemShut {NoStop}%
\bibitem [{\citenamefont {Guo}\ \emph {et~al.}(2015)\citenamefont {Guo},
  \citenamefont {Xu}, \citenamefont {Zou},\ and\ \citenamefont
  {Shao}}]{Guo2015}%
  \BibitemOpen
  \bibfield  {author} {\bibinfo {author} {\bibfnamefont {L.-S.}\ \bibnamefont
  {Guo}}, \bibinfo {author} {\bibfnamefont {B.-M.}\ \bibnamefont {Xu}},
  \bibinfo {author} {\bibfnamefont {J.}~\bibnamefont {Zou}}, \ and\ \bibinfo
  {author} {\bibfnamefont {B.}~\bibnamefont {Shao}},\ }\bibfield  {title}
  {\enquote {\bibinfo {title} {Improved thermometry of low-temperature quantum
  systems by a ring-structure probe},}\ }\href {\doibase
  10.1103/PhysRevA.92.052112} {\bibfield  {journal} {\bibinfo  {journal} {Phys.
  Rev. A}\ }\textbf {\bibinfo {volume} {92}},\ \bibinfo {pages} {052112}
  (\bibinfo {year} {2015})}\BibitemShut {NoStop}%
\bibitem [{\citenamefont {Mehboudi}\ \emph {et~al.}(2015)\citenamefont
  {Mehboudi}, \citenamefont {Moreno-Cardoner}, \citenamefont {De~Chiara},\ and\
  \citenamefont {Sanpera}}]{MehboudiNJP2015}%
  \BibitemOpen
  \bibfield  {author} {\bibinfo {author} {\bibfnamefont {M.}~\bibnamefont
  {Mehboudi}}, \bibinfo {author} {\bibfnamefont {M.}~\bibnamefont
  {Moreno-Cardoner}}, \bibinfo {author} {\bibfnamefont {G.}~\bibnamefont
  {De~Chiara}}, \ and\ \bibinfo {author} {\bibfnamefont {A.}~\bibnamefont
  {Sanpera}},\ }\bibfield  {title} {\enquote {\bibinfo {title} {Thermometry
  precision in strongly correlated ultracold lattice gases},}\ }\href {\doibase
  10.1088/1367-2630/17/5/055020} {\bibfield  {journal} {\bibinfo  {journal}
  {New J. Phys.}\ }\textbf {\bibinfo {volume} {17}},\ \bibinfo {pages} {055020}
  (\bibinfo {year} {2015})}\BibitemShut {NoStop}%
\bibitem [{\citenamefont {Mehboudi}\ \emph {et~al.}(2016)\citenamefont
  {Mehboudi}, \citenamefont {Correa},\ and\ \citenamefont
  {Sanpera}}]{MehboudiPRA2016}%
  \BibitemOpen
  \bibfield  {author} {\bibinfo {author} {\bibfnamefont {M.}~\bibnamefont
  {Mehboudi}}, \bibinfo {author} {\bibfnamefont {L.~A.}\ \bibnamefont
  {Correa}}, \ and\ \bibinfo {author} {\bibfnamefont {A.}~\bibnamefont
  {Sanpera}},\ }\bibfield  {title} {\enquote {\bibinfo {title} {Achieving
  sub-shot-noise sensing at finite temperatures},}\ }\href {\doibase
  10.1103/PhysRevA.94.042121} {\bibfield  {journal} {\bibinfo  {journal} {Phys.
  Rev. A}\ }\textbf {\bibinfo {volume} {94}},\ \bibinfo {pages} {042121}
  (\bibinfo {year} {2016})}\BibitemShut {NoStop}%
\bibitem [{\citenamefont {Paris}(2016)}]{ParisJPA2016}%
  \BibitemOpen
  \bibfield  {author} {\bibinfo {author} {\bibfnamefont {M.~G.~A.}\
  \bibnamefont {Paris}},\ }\bibfield  {title} {\enquote {\bibinfo {title}
  {Achieving the landau bound to precision of quantum thermometry in systems
  with vanishing gap},}\ }\href
  {http://iopscience.iop.org/article/10.1088/1751-8113/49/3/03LT02} {\bibfield
  {journal} {\bibinfo  {journal} {J. Phys. A: Math. Theor.}\ }\textbf {\bibinfo
  {volume} {49}},\ \bibinfo {pages} {03LT02} (\bibinfo {year}
  {2016})}\BibitemShut {NoStop}%
\bibitem [{\citenamefont {De~Pasquale}\ \emph {et~al.}(2016)\citenamefont
  {De~Pasquale}, \citenamefont {Rossini}, \citenamefont {Fazio},\ and\
  \citenamefont {Giovannetti}}]{DePasqualeNatComm2016}%
  \BibitemOpen
  \bibfield  {author} {\bibinfo {author} {\bibfnamefont {A.}~\bibnamefont
  {De~Pasquale}}, \bibinfo {author} {\bibfnamefont {D.}~\bibnamefont
  {Rossini}}, \bibinfo {author} {\bibfnamefont {R.}~\bibnamefont {Fazio}}, \
  and\ \bibinfo {author} {\bibfnamefont {V.}~\bibnamefont {Giovannetti}},\
  }\bibfield  {title} {\enquote {\bibinfo {title} {Local quantum
  thermometry},}\ }\href {http://www.nature.com/articles/ncomms12782}
  {\bibfield  {journal} {\bibinfo  {journal} {Nat. Commun.}\ }\textbf {\bibinfo
  {volume} {7}},\ \bibinfo {pages} {12782} (\bibinfo {year}
  {2016})}\BibitemShut {NoStop}%
\bibitem [{\citenamefont {De~Pasquale}\ \emph {et~al.}(2017)\citenamefont
  {De~Pasquale}, \citenamefont {Yuasa},\ and\ \citenamefont
  {Giovannetti}}]{DePasqualePRA2017}%
  \BibitemOpen
  \bibfield  {author} {\bibinfo {author} {\bibfnamefont {A.}~\bibnamefont
  {De~Pasquale}}, \bibinfo {author} {\bibfnamefont {K.}~\bibnamefont {Yuasa}},
  \ and\ \bibinfo {author} {\bibfnamefont {V.}~\bibnamefont {Giovannetti}},\
  }\bibfield  {title} {\enquote {\bibinfo {title} {Estimating temperature via
  sequential measurements},}\ }\href {\doibase 10.1103/PhysRevA.96.012316}
  {\bibfield  {journal} {\bibinfo  {journal} {Phys. Rev. A}\ }\textbf {\bibinfo
  {volume} {96}},\ \bibinfo {pages} {012316} (\bibinfo {year}
  {2017})}\BibitemShut {NoStop}%
\bibitem [{\citenamefont {De~Palma}\ \emph {et~al.}(2017)\citenamefont
  {De~Palma}, \citenamefont {De~Pasquale},\ and\ \citenamefont
  {Giovannetti}}]{DePasquale2017}%
  \BibitemOpen
  \bibfield  {author} {\bibinfo {author} {\bibfnamefont {G.}~\bibnamefont
  {De~Palma}}, \bibinfo {author} {\bibfnamefont {A.}~\bibnamefont
  {De~Pasquale}}, \ and\ \bibinfo {author} {\bibfnamefont {V.}~\bibnamefont
  {Giovannetti}},\ }\bibfield  {title} {\enquote {\bibinfo {title} {Universal
  locality of quantum thermal susceptibility},}\ }\href {\doibase
  10.1103/PhysRevA.95.052115} {\bibfield  {journal} {\bibinfo  {journal} {Phys.
  Rev. A}\ }\textbf {\bibinfo {volume} {95}},\ \bibinfo {pages} {052115}
  (\bibinfo {year} {2017})}\BibitemShut {NoStop}%
\bibitem [{\citenamefont {Campbell}\ \emph {et~al.}(2017)\citenamefont
  {Campbell}, \citenamefont {Mehboudi}, \citenamefont {De~Chiara},\ and\
  \citenamefont {Paternostro}}]{CampbellNJP2017}%
  \BibitemOpen
  \bibfield  {author} {\bibinfo {author} {\bibfnamefont {S.}~\bibnamefont
  {Campbell}}, \bibinfo {author} {\bibfnamefont {M.}~\bibnamefont {Mehboudi}},
  \bibinfo {author} {\bibfnamefont {G.}~\bibnamefont {De~Chiara}}, \ and\
  \bibinfo {author} {\bibfnamefont {M.}~\bibnamefont {Paternostro}},\
  }\bibfield  {title} {\enquote {\bibinfo {title} {Global and local thermometry
  schemes in coupled quantum systems},}\ }\href {\doibase
  10.1088/1367-2630/aa7fac} {\bibfield  {journal} {\bibinfo  {journal} {New J.
  Phys.}\ }\textbf {\bibinfo {volume} {19}},\ \bibinfo {pages} {103003}
  (\bibinfo {year} {2017})}\BibitemShut {NoStop}%
\bibitem [{\citenamefont {Mancino}\ \emph {et~al.}(2017)\citenamefont
  {Mancino}, \citenamefont {Sbroscia}, \citenamefont {Gianani}, \citenamefont
  {Roccia},\ and\ \citenamefont {Barbieri}}]{Mancino2017}%
  \BibitemOpen
  \bibfield  {author} {\bibinfo {author} {\bibfnamefont {L.}~\bibnamefont
  {Mancino}}, \bibinfo {author} {\bibfnamefont {M.}~\bibnamefont {Sbroscia}},
  \bibinfo {author} {\bibfnamefont {I.}~\bibnamefont {Gianani}}, \bibinfo
  {author} {\bibfnamefont {E.}~\bibnamefont {Roccia}}, \ and\ \bibinfo {author}
  {\bibfnamefont {M.}~\bibnamefont {Barbieri}},\ }\bibfield  {title} {\enquote
  {\bibinfo {title} {Quantum simulation of single-qubit thermometry using
  linear optics},}\ }\href {\doibase 10.1103/PhysRevLett.118.130502} {\bibfield
   {journal} {\bibinfo  {journal} {Phys. Rev. Lett.}\ }\textbf {\bibinfo
  {volume} {118}},\ \bibinfo {pages} {130502} (\bibinfo {year}
  {2017})}\BibitemShut {NoStop}%
\bibitem [{\citenamefont {Raitz}\ \emph {et~al.}(2015)\citenamefont {Raitz},
  \citenamefont {Souza}, \citenamefont {Auccaise}, \citenamefont {Sarthour},\
  and\ \citenamefont {Oliveira}}]{Oliveira2015}%
  \BibitemOpen
  \bibfield  {author} {\bibinfo {author} {\bibfnamefont {C.}~\bibnamefont
  {Raitz}}, \bibinfo {author} {\bibfnamefont {A.~M.}\ \bibnamefont {Souza}},
  \bibinfo {author} {\bibfnamefont {R.}~\bibnamefont {Auccaise}}, \bibinfo
  {author} {\bibfnamefont {R.~S.}\ \bibnamefont {Sarthour}}, \ and\ \bibinfo
  {author} {\bibfnamefont {I.~S.}\ \bibnamefont {Oliveira}},\ }\bibfield
  {title} {\enquote {\bibinfo {title} {Experimental implementation of a
  nonthermalizing quantum thermometer},}\ }\href {\doibase
  10.1007/s11128-014-0858-z} {\bibfield  {journal} {\bibinfo  {journal}
  {Quantum Inf. Process.}\ }\textbf {\bibinfo {volume} {14}},\ \bibinfo {pages}
  {37} (\bibinfo {year} {2015})}\BibitemShut {NoStop}%
\bibitem [{\citenamefont {Frey}(2016)}]{Frey2016QIP}%
  \BibitemOpen
  \bibfield  {author} {\bibinfo {author} {\bibfnamefont {M.~R.}\ \bibnamefont
  {Frey}},\ }\bibfield  {title} {\enquote {\bibinfo {title} {Quantum speed
  limits---primer, perspectives, and potential future directions},}\ }\href
  {\doibase 10.1007/s11128-016-1405-x} {\bibfield  {journal} {\bibinfo
  {journal} {Quantum Inf. Process.}\ }\textbf {\bibinfo {volume} {15}},\
  \bibinfo {pages} {3919--3950} (\bibinfo {year} {2016})}\BibitemShut {NoStop}%
\bibitem [{\citenamefont {Deffner}\ and\ \citenamefont
  {Campbell}(2017)}]{Deffner2017JPA}%
  \BibitemOpen
  \bibfield  {author} {\bibinfo {author} {\bibfnamefont {S.}~\bibnamefont
  {Deffner}}\ and\ \bibinfo {author} {\bibfnamefont {S.}~\bibnamefont
  {Campbell}},\ }\bibfield  {title} {\enquote {\bibinfo {title} {{Quantum speed
  limits: from Heisenberg's uncertainty principle to optimal quantum
  control}},}\ }\href {https://doi.org/10.1088/1751-8121/aa86c6} {\bibfield
  {journal} {\bibinfo  {journal} {J. Phys. A: Math. Theor.}\ }\textbf {\bibinfo
  {volume} {50}},\ \bibinfo {pages} {453001} (\bibinfo {year}
  {2017})}\BibitemShut {NoStop}%
\bibitem [{\citenamefont {Giovannetti}\ \emph {et~al.}(2006)\citenamefont
  {Giovannetti}, \citenamefont {Lloyd},\ and\ \citenamefont
  {Maccone}}]{LloydPRL2006}%
  \BibitemOpen
  \bibfield  {author} {\bibinfo {author} {\bibfnamefont {V.}~\bibnamefont
  {Giovannetti}}, \bibinfo {author} {\bibfnamefont {S.}~\bibnamefont {Lloyd}},
  \ and\ \bibinfo {author} {\bibfnamefont {L.}~\bibnamefont {Maccone}},\
  }\bibfield  {title} {\enquote {\bibinfo {title} {Quantum metrology},}\ }\href
  {\doibase 10.1103/PhysRevLett.96.010401} {\bibfield  {journal} {\bibinfo
  {journal} {Phys. Rev. Lett.}\ }\textbf {\bibinfo {volume} {96}},\ \bibinfo
  {pages} {010401} (\bibinfo {year} {2006})}\BibitemShut {NoStop}%
\bibitem [{\citenamefont {Giovannetti}\ \emph {et~al.}(2012)\citenamefont
  {Giovannetti}, \citenamefont {Lloyd},\ and\ \citenamefont
  {Maccone}}]{LloydPRL2012}%
  \BibitemOpen
  \bibfield  {author} {\bibinfo {author} {\bibfnamefont {V.}~\bibnamefont
  {Giovannetti}}, \bibinfo {author} {\bibfnamefont {S.}~\bibnamefont {Lloyd}},
  \ and\ \bibinfo {author} {\bibfnamefont {L.}~\bibnamefont {Maccone}},\
  }\bibfield  {title} {\enquote {\bibinfo {title} {Quantum measurement bounds
  beyond the uncertainty relations},}\ }\href {\doibase
  10.1103/PhysRevLett.108.260405} {\bibfield  {journal} {\bibinfo  {journal}
  {Phys. Rev. Lett.}\ }\textbf {\bibinfo {volume} {108}},\ \bibinfo {pages}
  {260405} (\bibinfo {year} {2012})}\BibitemShut {NoStop}%
\bibitem [{\citenamefont {Nieto}(1995)}]{Nieto1995}%
  \BibitemOpen
  \bibfield  {author} {\bibinfo {author} {\bibfnamefont {A.}~\bibnamefont
  {Nieto}},\ }\bibfield  {title} {\enquote {\bibinfo {title} {Evaluating sums
  over the matsubara frequencies},}\ }\href {\doibase
  http://dx.doi.org/10.1016/0010-4655(95)00061-J} {\bibfield  {journal}
  {\bibinfo  {journal} {Computer Physics Communications}\ }\textbf {\bibinfo
  {volume} {92}},\ \bibinfo {pages} {54 -- 64} (\bibinfo {year}
  {1995})}\BibitemShut {NoStop}%
\bibitem [{\citenamefont {Zanardi}\ \emph {et~al.}(2008)\citenamefont
  {Zanardi}, \citenamefont {Paris},\ and\ \citenamefont
  {Campos~Venuti}}]{ZanardiPRA2008}%
  \BibitemOpen
  \bibfield  {author} {\bibinfo {author} {\bibfnamefont {P.}~\bibnamefont
  {Zanardi}}, \bibinfo {author} {\bibfnamefont {M.~G.~A.}\ \bibnamefont
  {Paris}}, \ and\ \bibinfo {author} {\bibfnamefont {L.}~\bibnamefont
  {Campos~Venuti}},\ }\bibfield  {title} {\enquote {\bibinfo {title} {Quantum
  criticality as a resource for quantum estimation},}\ }\href {\doibase
  10.1103/PhysRevA.78.042105} {\bibfield  {journal} {\bibinfo  {journal} {Phys.
  Rev. A}\ }\textbf {\bibinfo {volume} {78}},\ \bibinfo {pages} {042105}
  (\bibinfo {year} {2008})}\BibitemShut {NoStop}%
\bibitem [{\citenamefont {Husimi}(1953)}]{Husimi1953}%
  \BibitemOpen
  \bibfield  {author} {\bibinfo {author} {\bibfnamefont {K}~\bibnamefont
  {Husimi}},\ }\bibfield  {title} {\enquote {\bibinfo {title} {{Miscellanea in
  elementary quantum mechanics, II}},}\ }\href
  {papers://eb31b660-126c-4e97-a4f0-645efff98fb4/Paper/p198} {\bibfield
  {journal} {\bibinfo  {journal} {Prog. Theor. Phys.}\ }\textbf {\bibinfo
  {volume} {9}},\ \bibinfo {pages} {381} (\bibinfo {year} {1953})}\BibitemShut
  {NoStop}%
\bibitem [{\citenamefont {Taddei}\ \emph {et~al.}(2013)\citenamefont {Taddei},
  \citenamefont {Escher}, \citenamefont {Davidovich},\ and\ \citenamefont
  {de~Matos~Filho}}]{TaddeiPRL}%
  \BibitemOpen
  \bibfield  {author} {\bibinfo {author} {\bibfnamefont {M.~M.}\ \bibnamefont
  {Taddei}}, \bibinfo {author} {\bibfnamefont {B.~M.}\ \bibnamefont {Escher}},
  \bibinfo {author} {\bibfnamefont {L.}~\bibnamefont {Davidovich}}, \ and\
  \bibinfo {author} {\bibfnamefont {R.~L.}\ \bibnamefont {de~Matos~Filho}},\
  }\bibfield  {title} {\enquote {\bibinfo {title} {Quantum speed limit for
  physical processes},}\ }\href {\doibase 10.1103/PhysRevLett.110.050402}
  {\bibfield  {journal} {\bibinfo  {journal} {Phys. Rev. Lett.}\ }\textbf
  {\bibinfo {volume} {110}},\ \bibinfo {pages} {050402} (\bibinfo {year}
  {2013})}\BibitemShut {NoStop}%
\bibitem [{\citenamefont {Deffner}\ and\ \citenamefont
  {Lutz}(2013)}]{DeffnerPRL}%
  \BibitemOpen
  \bibfield  {author} {\bibinfo {author} {\bibfnamefont {S.}~\bibnamefont
  {Deffner}}\ and\ \bibinfo {author} {\bibfnamefont {E.}~\bibnamefont {Lutz}},\
  }\bibfield  {title} {\enquote {\bibinfo {title} {Quantum speed limit for
  non-markovian dynamics},}\ }\href {\doibase 10.1103/PhysRevLett.111.010402}
  {\bibfield  {journal} {\bibinfo  {journal} {Phys. Rev. Lett.}\ }\textbf
  {\bibinfo {volume} {111}},\ \bibinfo {pages} {010402} (\bibinfo {year}
  {2013})}\BibitemShut {NoStop}%
\end{thebibliography}%

\end{document}